\documentstyle[12pt]{article}
\newcommand{\be}{\nopagebreak\begin{equation}}
\newcommand{\ee}{\end{equation}}
\newcommand{\ba}{\begin{array}}
\newcommand{\ea}{\end{array}}
\newcommand{\bp}{\begin{picture}}
\newcommand{\ep}{\end{picture}}
\newcommand{\eol}{\\[1pc]}
\newcommand{\ftnt}[1]{$\mbox{}^($\footnote{$\mbox{}^)$ #1}$\mbox{}^)$}
\newcommand{\rf}[1]{Ref.~\cite{#1}}
\newcommand{\eq}[1]{Eq.~(\ref{#1})}
\newcommand{\bi}[6]{\bibitem{#1}{#2, }{\sl #3 }{\bf #4}{ (#5)}{ #6}}
\newcommand{\wt}[1]{\widetilde{#1}}
\newcommand{\wh}[1]{\widehat{#1}}
\newcommand{\ie}{{\it i.e., }}
\newcommand{\eg}{{\it e.g., }}
\renewcommand{\d}{\partial}
\newcommand{\tr}{{\rm tr}\;}
\newcommand{\inv}{^{\scs -1}}
\newcommand{\tsp}{^{\scs{\rm T}}}
\newcommand{\scr}{\scriptstyle}
\newcommand{\scs}{\scriptscriptstyle}
\newcommand{\nnl}{\noindent}

\newcommand{\mv}[1]{\langle #1 \rangle}

\newcommand{\newsection}{    
\setcounter{equation}{0}
\section}

\begin{document}
\begin{titlepage}
\begin{flushright}
Preprint ITEP-TH-13/98\\
March 1998\\
\end{flushright}
\vspace*{5pc}
\begin{center}
{\huge \bf Three-dimensional simplicial gravity and combinatorics of group
presentations}
\end{center}
\vspace{2pc}
\begin{center}
 {\Large D.V. Boulatov}\\
\vspace{1pc}
{\em Institute for Theoretical and Experimental Physics (ITEP)\\
B.Cheremushkinskaya 25, Moscow, Russia}\\boulatov@heron.itep.ru
\vspace{2pc}
\end{center}
\begin{center}
{\large\bf Abstract}
\end{center}

\nnl We demonstrate how some problems arising in simplicial quantum
gravity can be successfully addressed within the framework of combinatorial
group theory. In particular, we argue that the number of simplicial
3-manifolds having a fixed homology type grows exponentially with the number
of tetrahedra they are made of. We propose a model of 3D gravity
interacting with scalar fermions, some restriction of which gives the
2-dimensional self-avoiding-loop-gas matrix model. We propose a qualitative
picture of the phase structure of 3D simplicial gravity compatible with the
numerical experiments and available analytical results.

\vfill
\end{titlepage}     

\newsection{Introduction}

The success of the matrix models as a theory of 2D quantum gravity and 
non-critical strings has brought about a hope that analogous discrete 
approaches might be instructive in higher dimensions as well. The most 
natural way to introduce discretized quantum gravity is to consider 
simplicial complexes instead of continuous manifolds. Then, a path 
integral over metrics (which is the crux of any approach to quantum 
gravity) can be simply defined as a sum over all complexes having some 
fixed properties. For example, it is natural to restrict their topology. 
In the present paper, we consider only the 3-dimensional case, where, 
by definition, a simplicial complex is a collection of tetrahedra glued 
along their faces in such a way that any two of them can have at most one 
triangle in common.

To introduce metric properties, one can assume that complexes represent
piece-wise linear manifolds glued of equilateral tetrahedra \cite{Regge}.  The
volume is proportional to the number of them.  In the continuum limit, as this
number grows, the edge length, $a$, simultaneously tends to zero: $a\to0$. In
three dimensions, there is a one-to-one correspondence between classes of
piece-wise linear and smooth manifolds. In other words, any homeomorphism can
equally be approximated by either a piece-wise linear or smooth map.
This property makes the model self-consistent.

Within a piece-wise linear approximation, the curvature is singular: the 
space is flat everywhere off edges. Therefore, one should consider 
integrated quantities, \eg the mean scalar curvature

\be \int d^3x\; \sqrt g R = a\Big(2\pi N_1 -6N_3\arccos \frac13\Big) \ee
where $a$ is the lattice spacing and $N_0,\ N_1,\ N_2,\ N_3$ are the numbers
of vertices, edges, triangles and tetrahedra, respectively.

For manifolds in 3 dimensions, the Euler character vanishes $\chi =
N_0-N_1+N_2-N_3=0$.  Together with the other constraint $N_2=2N_3$, it means
that only 2 of the $N_i$'s are independent, $N_1=N_0+N_3$, and a natural
lattice action depends on 2 dimensionless parameters. Now, we are in a
position to define the simplicial-gravity partition function
\cite{ADJ,3dsim,3dsim2,B}:

\be
{\bf Z}(\alpha,\mu)=\sum_{\{C\}} e^{\alpha N_0 - \mu N_3}
=\sum_{N_3} Z_{N_3}(\alpha) e^{-\mu N_3}
\label{Z}
\ee 
where $\sum_{\{C\}}$ is a sum over some class of complexes (\eg of a
fixed topology).

It can be shown that, if $N_3$ is fixed, $N_0$ is restricted from above as
$N_0<\frac13N_3+const$. Therefore, it is natural, keeping $\alpha$ fixed,
to tend $\mu$ to its critical value, $\mu_c$, at which the sum over $N_3$
in \eq{Z} becomes divergent. In the vicinity of $\mu_c$ one could expect
to find critical behavior corresponding to a continuum limit of the model (of
course, it can require a fine-tuning of the second parameter $\alpha$).

For the model to be physically consistent, the free energy per unit volume
$\frac1{N_3}\log Z_{N_3}(\alpha)$ has to be finite, which means that the
series over $N_3$ in \eq{Z} has to have a finite radius of convergence:
$0<\mu_c<+\infty$. If it holds, statistical models defined on such an
ensemble of fluctuating lattices possess reasonable thermodynamical
behavior. Thus one can hope to describe matter coupled to quantum
gravity. This program has been successfully carried out in two dimensions
with help of the matrix model technique \cite{KKM,O(n)}. In the 3D
case, analytic tools are lacking and numerical simulations have so far
been the only source of information \cite{3dsim,3dsim2}.

Many questions, which are simple in two dimensions, become stubborn
 in three. For example, 2-manifolds are classified according to
values of the Euler character, therefore topology, homotopy and homology are
actually equivalent descriptions. In three dimensions, sets of manifolds
characterized according to these 3 criteria are essentially different. A
natural question to ask is how large a class of complexes $\{C\}$ in
\eq{Z} can be. We shall argue that it is sufficient to fixe only the first
homology group $H_1(C,Z)$ in order to have a finite value of $\mu_c$. It
can be reformulated as the following statement: ``The number of simplicial
manifolds constructed of the given number of tetrahedra, $N_3$, and having
a fixed homology type ({\em viz.,} the first Betti number and the torsion
coefficients) grows at most exponentially as a function of $N_3$''. 

Let us notice that the existence of the exponential bound 
is not broken down, if one weights every complex with a positive 
weight growing at most exponentially with $N_3$. 
As well, if one manages to represent a sum over complexes as a weighted 
sum over another class of objects and to prove that the weights are 
exponentially bounded, then one can simply
estimate the number of objects in this new class ignoring the weights.

The simplest example of a weighted sum of complexes is given by the
canonical partition function $Z_{N_3}(\alpha)$ introduced in
\eq{Z}. Another example of physical interest is the partition function in
the presence of free matter fields. To introduce it let us consider the
set of polyhedral complexes dual to simplicial ones. Their 1-skeletons are some
4-valent graphs, $G^{(4)}$, whose vertices correspond to tetrahedra, and
links are dual to triangles. Such a graph can be defined by the
adjacency matrix

\be 
G^{(4)}_{ij}=\left\{\ba{ll} 
1 & \mbox{if vertices $i$ and $j$ are connected by a link}\\ 
0 & \mbox{otherwise} 
\ea \right.  
\ee

Of course, the set of 4-valent graphs is not identical to the totality of
simplicial complexes. Similarly, in two dimensions, the set of ordinary
$\phi^3$ Feynmann diagrams is different from that of triangulations --
one has to introduce the notion of fat graphs to establish the
equivalence.  

If one attaches the $n$-component vector $x^{\mu}_i\ 
(\mu=1,\ldots,n)$ to the $i$-th vertex, the corresponding gaussian integral 
can be performed explicitly for every given graph $G^{(4)}$:

\be
Z_{matter}(G^{(4)})=
\int \prod_{i=1}^{N_3-1}\prod_{\mu=1}^n d x^{\mu}_i \exp\Big[ -\frac12 
\sum_{i,j=1}^{N_3-1}G^{(4)}_{ij}(x^{\nu}_i-x^{\nu}_j)^2\Big] =
\Big(\det\mbox{}' L^{(4)}\Big)^{-\frac{n}2}
\label{gint}
\ee
where the discrete Laplacian is given by

\be
L^{(4)}_{ij}=4\delta_{ij}-G^{(4)}_{ij}
\ee
There is the nice combinatorial representation of the determinant 
given by Kirchhoff's theorem \cite{Biggs}:

\be
\det\mbox{}' L = |T(G)|
\ee
where $T(G)$ is the set of all possible spanning trees embedded into
a graph $G$ or, equivalently, a number of connected trees which can be
obtained from the graph by cutting its links.  To
kill the zero mode in \eq{gint}, we fixed the field at the $N_3$'th
vertex. The theorem states that $\det\mbox{}' L$ does not depend on a choice of
the vertex. Therefore, $|T(G)|$ is the number of rooted trees.

The number of spanning trees of a $k$-valent graph $G^{(k)}_n$ with $n$
vertices can be estimated from above as \cite{Biggs}

\be
|T(G^{(k)}_n)|\leq \frac1n\Bigg(\frac{nk}{n-1}\Bigg)^{n-1}
\ee

The case we are particularly interested in is the 2-component Grassmann
field, where we obtain the gravity+matter partition function in the form

\be
{\bf Z}^{(1)}(\alpha,\mu)=\sum_{\{C\}}\det\mbox{}' L\; 
e^{\alpha N_0 - \mu N_3}
=\sum_{N_3} Z^{(1)}_{N_3}(\alpha)e^{-\mu N_3}
\label{Z1a}
\ee
The weights, $ \det\mbox{}' L$, are positive integers, hence,

\be
Z_{N_3}(\alpha)<Z^{(1)}_{N_3}(\alpha)
\ee

In two dimensions, this type of matter has the central charge $c=-2$ and the 
corresponding matrix model has been solved explicitly by purely
combinatorial means \cite{KKM}. 
We can repeat the same trick in 3 dimensions:

\be 
{\bf Z}^{(1)}(\alpha,\mu)=\sum_{\{C\}}\det\mbox{}' L\; e^{\alpha N_0-\mu
N_3} =\sum_{\{C\}}\sum_{\{\wt{T}(C)\}}e^{\alpha N_0-\mu N_3}=\sum_{\{\wt{T}\}}
\sum_{\{C(\wt{T})\}}e^{\alpha N_0-\mu N_3} 
\label{Z1}
\ee 

\nnl Namely, we have the sum over complexes, $\sum_{\{C\}}$, 
weighted with the number of spanning trees in
the dual 1-skeleton of each complex, $\sum_{\{\wt{T}(C)\}}$. 
We can obtain exactly the
same configurations by taking the sum over all possible trees,
$\sum_{\{\wt{T}\}}$, weighted with the number of complexes, 
$\sum_{\{C(\wt{T})\}}$,
which can be recovered from a given tree $\wt{T}$ by restoring cut links.

In three dimensions, the appearing trees correspond to spherical
simplicial balls obtained starting with a single tetrahedron
by subsequently gluing other tetrahedra to faces of the boundary.
If we denote as $n_0,\ n_1$ and $n_2$ numbers of vertices, edges and 
triangles on the boundary of a ball, we find that 

\be
n_0=N_3+3 \hspace{2pc} n_2=2(N_3+1) 
\ee

In contrast to the 2-dimensional case, it is extremely difficult in three
dimensions to control a topology of complexes constructed from a ball by
pairwise identifications of triangles belonging to the boundary. Therefore,
we shall not address the homeomorphic properties of complexes in this
paper. On the other hand, homotopy information is nicely coded in the
fundamental group, which is going to be the basic object for us. The
combinatorial group theory provides us with the most adequate mathematical
framework within which all physical questions can be asked and, hopefully,
answered.

The outline of the paper is the following.\\
In Section~2 we collect some basic definitions and results from
combinatorial group theory and 3-dimensional topology.\\
In Section~3 we give an interpretation of the 3D gravity partition function
in terms of group presentations.\\
In Section~4 we argue that the number of simplicial complexes of a given
volume having fixed homology type grows exponentially as a function of the
volume. \\
In Section~5 we introduce a reduced model and establish a connection with
the 2D loop gas matrix model.\\
Section~6 is devoted to a discussion.

The present paper can be regarded as an extended version of the previous
one \cite{Bo}. A reader must be aware that Ref.~\cite{Bo} contains some
incorrect statements. In particular, the proof of the main theorem stated
in it is not mathematically satisfactory.


\newsection{Complexes and group presentations}

Although almost all the mathematical material collected in this section can be
found in the standard reference books on lower dimensional topology and
combinatorial group theory (see for example \cite{Still,John}), it is
convenient to repeat it here for reader's conveninience. Let us start with
some definitions. 

If it is not specifically stated, by a complex we always mean a finite cell
(polyhedral) decomposition of a compact closed orientable 3-manifold. Any
complex, $C$, can be constructed by first assembling the 1-dimensional cells
(giving the 1-skeleton, $K_1$), then attaching 2-cells (giving the
2-skeleton, $K_2$) and analogously for $K_3$. 

A simplicial complex is a complex in which $k$-dimensional cells are
$k$-simplexes, \ie points, segments, triangles and tetrahedra. The intersection
of any two simplexes either is empty or consists of exactly one less
dimensional simplex. Any simplex in a simplicial complex can be
unambiguously determined, including an orientation, by a list of vertices
from its boundary. It means, in particular, that any edge connects a
pair of distinct vertices.

The fundamental group $\pi_1(C)$ is determined by the 2-skeleton of a
complex: $\pi_1(C)\equiv\pi_1(K_2)$. The standard algorithm to read off a
presentation $\pi_1(C)$ is the following:
\begin{enumerate}
\item Construct a spanning tree $T$ of the 1-skeleton of $C$.
\item Put into correspondence a generator, $a_i$, to every link of $K_1$
which is not in $T$ (\ie to each cut link). Links belonging to $T$ are
formally associated with the identity,~1.
\item Fixe some orientation of the links of $K_1$; a change of the
orientation corresponds to the inversion of the generator: $a_i\to
a_i^{-1}$. 
\item For every face in the 2-skeleton, the cyclic order of the oriented
links forming its perimeter gives a relation: $r_j=a_{i_1}^{\epsilon_1}
a_{i_2}^{\epsilon_2}\ldots a_{i_\ell}^{\epsilon_\ell}=1$; $\epsilon_i=\pm1$
according to the orientation of the $i$'th link (one puts the
identity, 1, on places corresponding to the links which are in $T$). 
\end{enumerate}
As any spanning tree cuts all loops in $K_1$, all relations obtained in
this way are non-trivial (the trivial relation is, by definition, of the
form 1=1).

In this way one finds the fundamental group presentation

\be
\pi_1(C)=\mv{a_1a_2\ldots a_n|r_1r_2\ldots r_m}
\label{present}
\ee
where $m$ equals the number of 2-cells in $C$ and $n$ is not bigger that
the number of 1-cells.

This procedure gives a nonreduced group presentation in the sense that
presentations differing by the trivial cancellations, $aa\inv\to1$ and
$a\inv a\to1$, correspond to distinct complexes. The simplest example is
the following identical presentations of the trivial group

\be
\begin{array}{l}
P'_E=\mv{a|a,a,a}\\
P''_E=\mv{a|aa\inv a}
\end{array}
\ee 

In both cases $K_2$ consists of the single triangle but, in
the second case, its edges are identified in such a way that the
corresponding complex has only one 0-cell and only one 1-cell. The
2-skeleton here is the so-called Zeeman's 
``Dunce hat''\ftnt{I am indebted to J.Howie for pointing me out this
example.}, a closed contractible fake-surface. Both
3-complexes represent $S^3$ \cite{Zeeman}.

Therefore, in order to maintain correspondence between presentations and
complexes, one has to consider nonreduced presentations. However, as we
shall see later, in the case of simplicial 3-complexes, this subtlety can
easily be overcome. Therefore, in what follows, we shall use the term
``presentation'' hoping it will always be clear from a context what is
meant in every particular case.

Two presentations describe the same complex, if they can be
transformed into each other by a sequence of the following transformations:

\begin{enumerate}
\item permutation of generators or relators;
\item inversion of a generator, $a_i\to a\inv_i$, or a relator,
$r=a_1a_2\ldots a_k\to r\inv=a\inv_k\ldots a\inv_2a\inv_1$;
\item cyclic shift of a relator: 
$r=a_1a_2\ldots a_k\to a\inv_1ra_1=a_2\ldots a_k a_1$;
\end{enumerate}
These moves obviously divide the totality of all presentations into equivalence
classes. We shall always identify presentations from the same class.

It is known that all 3-manifolds possess finitely presentable fundamental
groups. Therefore, any of them can be described as the factor group

\be
G=F(X)/N
\ee
where $F(X)$ is the free group on $n$ generators, $X\equiv(a_1a_2\ldots a_n)$,
and $N$ is
the normal closure in $F(X)$ of a subgroup generated by some finite set of
words $R\subset F(X)$. The words constituting $R$ are the relators and this
is an equivalent description of the presentation. R form a basis of the
subgroup $\mv{R}$. Any other set of generators of $\mv{R}$ can be obtained
from $R$ by applying the Nielsen transformations:

\begin{itemize}
\item[N1)] replace $r_j$ by $r\inv_j$.
\item[N2)] replace $r_j$ by $r_jr_i$ with $i\neq j$.
\end{itemize}
This pair of transformations generate a group including permutations as a
subgroup. Any sequence of the Nielsen transformations carries a
presentation $\mv{X|R}$ into an equivalent one. Let us notice that they do
not change the number of generators.

To obtain the normal closure of $\mv{R}$, one has to add the third move,
namely, the conjugation by an arbitrary word $w\in F(X)$:\\

\noindent
N3)~replace $r_j$ by $wr_jw\inv$.\\

In order to establish the isomorphism of two groups, one has to use the
Tietze transformations:

\begin{itemize}
\item[T1)] add or remove a relator which is a consequence of other relators:
\[  \mv{X|R} \longleftrightarrow \mv{X|R,y} \]
where $y\in N$, N is the normal closure of $\mv{R}$ in F(X).
\item[T2)] add or remove a generator which is a word in the others:
\[  \mv{X|R} \longleftrightarrow \mv{X,x|R,x\inv w} \]
where $R$ and $w$ do not involve $x$; $w$ is arbitrary. This move is called
the prolongation.
\end{itemize}
It can be shown that these moves connect any two presentations of the same
group. It should be said that there is no correspondence between the
Tietze moves and the transformations of complexes: in most cases they
yield presentations having no realization in the form of 3-complexes.

One of the most profound connections between homotopy theory and group
presentations is established via the so-called formal defomations of
complexes \cite{Whit}. Let us introduce this notion. An elementary
$n$-expansion, $C'\nearrow C''$, is an attachment to $C'$ of an $n$-ball
$B_n$ along all of the boundary $\d B_n$ except one $(n-1)$-cell:
$C''=C'\bigcup B_n$, $C'\bigcap B_n\cong B_{n-1}$. If the complexes in
question are simplicial, the ball is simply the single tetrahedron. An
elementary $n$-collapse, $C'\searrow C''$, is the inverse of the elementary
$n$-expansion. A formal $n$-deformation, $C'\stackrel{n}{\nearrow\!\searrow}
C''$, is a finite sequence of the elementary expansions and collapses in
which the maximum dimension of the balls equals $n$.

If a complex $C$ can be $n$-deformed to a point, $*$, one says that $C$ is
$n$-collapsible. This term should not be confused with contractability,
which stays for the simply-connected complexes, $\pi_1(C)=1$. 

As was proven by P.Wright \cite{Wright}, any two contractible
2-complexes, $\pi_1(K'_2)=\pi_1(K''_2)=1$, can be transformed into each other
by a finite sequence of the formal deformations not exceeding the dimension 3,
$K'_2\stackrel{3}{\nearrow\!\searrow}K''_2$, if and only if the
corresponding presentations read off in the standard way from the two
2-complexes $K'_2$ and $K''_2$ can be transformed into each other by a
finite sequence of the Nielsen transformations extended by the trivial
prolongation. We shall call this set of moves the AC
transformations:

\begin{itemize}
\item[AC1)] cancellation of $aa\inv$ and its inverse.
\item[AC2)] $r_i\to r\inv_i$ and $r_i\to r_ir_j$ with $i\neq j$.
\item[AC3)] $r_i\to wr_iw\inv$ where $w\in F(X)$.
\item[AC4)] $\mv{X|R}\longleftrightarrow\mv{X,a|R,a}$ where $a\not\in X$, 
$a\not\in R$. 
\end{itemize}
These transformations will play an important role in the sequel.

It is not known whether arbitrary two 2-complexes having the same
 fundamental group can always be connected by a chain of the formal
 3-deformations, in other words, whether in three dimensions the simple
 homotopy type and the 3-deformation type coincide.  The famous
 Andrews-Curtis conjecture reads: {\em for any contractible compact $K_2$,
 $K_2\stackrel{3}{\nearrow\!\searrow}*$ holds}. It seems that this
 conjecture is commonly believed to be false \cite{Metz}.  In any case, as
 we have decided to forget about the notion of homeomorphy, the
 3-deformation classes of 2-spines of simplicial 3-manifolds seem to be the
 most natural finest classification in our context.

Another theorem of P.Wright \cite{Wright} says that
$K'_2\stackrel{2}{\nearrow\!\searrow}K''_2$ holds if and only if the
corresponding presentations can be transformed into each other by a
sequence of arbitrary prolongations (T2).  It is
important that the cancellation of $aa\inv$ is not allowed as an
independent move. For example, Zeeman's ``Dunce hat'' is not 2-collapsible
albeit contractible.

It is known that the first homology group is the abelianized
fundamental group: $H_1(C,Z)=\pi_1(C)/[\pi_1(C),\pi_1(C)]$. In the abelian
case, it is convenient to represent the defining relations in the form of a
linear system of equations:

\be
\wh{P}=\mv{x_1\ldots x_n|\sum_{j=1}^n \wh{R}_{ij}x_j}
\ee
where the relation matrix $\wh{R}$ has integer entries.

Another set of free generators can be obtained by rotating the vector
$(x_1\ldots x_n)$ by an invertible unimodular integral matrix:

\be
x'_i=\sum_{j=1}^n M_{ij} x_j \hspace{2pc} M\in SL(n,Z)
\ee
and analogously for relators.

It is known that pre- and post-multiplications by $SL(n,Z)$ matrices can be
expanded into sequences of the following elementary operations

\begin{enumerate}
\item permutation of rows.
\item multiplication of all elements in a row by $-1$.
\item addition of a row to another row.
\end{enumerate}
and the same for columns.

These operations correspond to the Nielsen transformations and give
a homomorphism of the group of the Nielsen moves into $SL(n,Z)$. The kernel
is generated by the conjugations, $r_j\to r_i r_j r\inv_i$.

An arbitrary rectangular $n\times m$ matrix can be transformed into the
diagonal form

\be
D=diag(d_1,d_2,\ldots,d_k) \hspace{2pc} k={\rm min}(n,m)
\ee
where $d_i$ are non-negative integers and $d_i$ divides $d_{i+1}$
$(d_i|d_{i+1})$. 

Thus, any abelian group has the unique presentation of the form

\be
\wh{G} = \mv{x_1\ldots x_n|x_1^{d_1},x_2^{d_2},\ldots,x_q^{d_q},C}
\cong Z_{p_1}\times Z_{p_2}\times\ldots Z_{p_q}\times
\underbrace{Z\times\ldots\times Z}_{b_1-times}
\ee
where $C=\{x_ix_jx\inv_ix\inv_j\}$ is the commutator subgroup of $F(X)$;
$Z_p$ is the cyclic group $\mv{x|x^p}$; $Z$ is the infinite cyclic group
$\mv{x|\ }$; $p_1>1;$ $p_q\neq0$; $p_i|p_{i+1}$. The number of $Z$ factors
is called the first Betti number, $b_1$, and the integers
$(p_1,p_2,\ldots,p_q)$ are the torsion coefficients. 

If $\wh{R}$ is a square matrix such that $\det \wh{R}\neq0$, then it
determines a finite abelian group $\wh{G}$ of the order
$|\wh{G}|=\pm\det\wh{R}$.  

It is well known that not all groups can be fundamental groups of
3-manifolds and not all presentations of a given group can be realized in
the form of a 2-skeleton of some 3-complex. The obvious (but not the only
one) source of the restrictions is the Poincar\'{e} duality. Given a
3-complex $C$, the dual complex $\wt{C}$ is constructed by putting
$k$-dimensional cells of $C$ into correspondence to $(3-k)$-cells of
$\wt{C}$. A presentation of $\pi(\wt{C})$ can be read off from the
2-skeleton, $\wt{K}_2$, of $\wt{C}$. As we consider manifolds,
$\pi_1(\wt{C})\cong\pi_1(C)$. It implies a kind of duality between
generators and relators in presentations of the fundamental groups of
3-manifolds.

Let us fixe a spanning tree $\wt{T}$ of the dual 1-skeleton,
$\wt{K}_1$. Relators in \eq{present} are in correspondence with links of
$\wt{K}_1$. This set of them is obviously excessive. A minimal set is given by
links which is not in $\wt{T}\in\wt{K}_1$. In this case, relators of
$\pi_1(C)$ correspond to generators of $\pi_1(\wt{C})$ and {\em vice versa}.
In particular, their numbers equal each other. Such presentations are
called balanced.


\newsection{Combinatorial meaning of the partition function}

Now we are in a position to interpret the partition function ${\bf
Z}^{(1)}(\alpha,\mu)$ introduced in \eq{Z1} in terms of combinatorial group
theory. Every simplicial ball from $\sum_{\{\wt{T}\}}$ determines a set of
$N_3+1$ generators of $\pi_1(\wt{C})$. Any 3-manyfold $M_3$ can be obtained
out of some ball by gluing its boundary triangles pairwise. This process
induces corresponding identifications of edges and vertices. The glued
triangles form a 2-dimensional spine\ftnt{By definition, a spine is a
2-skeleton in a cell decomposition with a single 3-cell.}, $K_2\subset M_3$,
having the Euler character $\chi(K_2)=1$. Every edge in $K_2$ gives a
relator of the fundamental group, $\pi_1(M_3)$, determined by a cyclic order
of oriented triangles attached to it. Thus we find

\be
{\bf Z}^{(1)}_{\pi_1}(\alpha,\mu)=\sum_{N_3,N_1}
\sum_{\{P_{N_3+1,N_1}\}} e^{\alpha N_1-(\alpha+\mu)N_3}
\ee

\nnl where $\sum_{\{{\scs P_{N_3+1,N_1}}\}}$ is the sum over all nonreduced
presentations on $N_3+1$ generators and $N_1$ relators associated with the
2-spines of simplicial complexes from a given class ($[\pi_1] = \{C:\
\pi_1(C)\ {\rm fixed}\}$):

\be
P_{n,m}\equiv\mv{X_n|R_m}=\mv{a_1a_2\ldots a_n|r_1r_2\ldots r_m}\in [\pi_1]
\ee

\nnl
where $r_i$'s are nonreduced words in $a_j^{\scs \pm1}$.  Every generator
$a_i\in X_n$ appeares exactly 3 times in $R_m$, simply because the triangle
has 3 edges.

Any spanning tree $T$ of the 1-skeleton $K_1$ of the 2-spine $K_2$ ($T\subset
K_1\subset K_2$) fixes a minimal subset of the relators $R'_{N_3+1} \subset
R_{N_1}$. The deformation retract of a tree is a point: $T\searrow *$, and
the fundamental group of any 3-manifold possesses a balanced
presentation.

It is convenient to weight every complex with the number of spanning
trees $T\subset K_1\subset K_2$. It can be done by introducing another
system of free matter fields attached to vertices of triangulations. Then,
analogously to \eq{Z1}, we define the partition function

\be 
{\bf Z}^{(2)}(\alpha,\mu)=\sum_{\{C\}} \det\mbox{}'L\;
\det\mbox{}'\wt{L} \;e^{\alpha N_0-\mu N_3}= \sum_{N_3} e^{-\mu
N_3}\sum_{\{H_{N_3}\in [\pi_1]\}} {\bf \Upsilon}(H_{N_3},\alpha)
\label{Z2}
\ee 

\nnl where $\sum_{\{H_{N_3}\in[\pi_1]\}}$ goes over all balanced
presentations from a given homotopy class:
$\{H_{N_3}\equiv\mv{X_{N_3+1}|R'_{N_3+1}}\}= [\pi_1]$. All such
presentations can be enumerated. Then, there exists an algorithm
\cite{Neuw} allowing, in principle, for determining if a corresponding
2-complex is a spine of some 3-manifold.

\be 
{\bf \Upsilon}(H_{N_3},\alpha) = \sum_{N_0} 
\sum_{\{P_{N_3+1,N_3+N_0}\cong H_{N_3}\}} e^{\alpha N_0} = 
\sum_{N_0} e^{\alpha N_0} \Upsilon_{N_0}(H_{N_3}) 
\label{Z2b}
\ee 

\nnl where $\sum_{\{P_{N_3+1,N_3+N_0}\cong H_{N_3}\}}$ is the sum over all
nonreduced presentations $\mv{X_{N_3+1}|R_{N_3+N_0}}$ which can be deduced
from a given balanced presentation $H_{N_3}=\mv{X_{N_3+1}|R_{N_3+1}}$
following a pattern of a spanning tree in some simplicial complex. The last
equality is a definition of $\Upsilon_{N_0}(H_{N_3})$.

In this way we divide the counting problem into two steps: we estimate
first the number of the balanced presentations in a class $\{H_{N_3}\}$ and
then the number of simplicial complexes giving any given balanced
presentation from $\{H_{N_3}\}$. The last number,
$\Upsilon_{N_0}(H_{N_3})$, is exponentially bounded because any
configuration has the spanning tree structure and the bound is provided by
the number of corresponding trees. To estimate the first sum, $|H_{N_3}\in\
[\pi_1]|$, is a fundamentally difficult combinatorial problem.

The balanced presentations enjoy the duality property. Therefore
$\sum_{\{H_{N_3}\}}$ can be equivalently described as the sum over the balanced
presentations in which the maximal length of relators does not exceed 3:
$\ell(r_j)\leq3,\ \forall j$. In this dual description, the cyllables
$a_ia\inv_i$ and $a^2_i$ never appear in the relators, because in that
case there would exist a circularly closed edge, which is
forbidden by the definition of the simplicial complex. Therefore we can
consider reduced relators without loss of generality.


\newsection{Estimates for abelian presentations}

Let us formulate the following simple fact as

\noindent
{\sc Proposition~1.} If the maximum length of relators equals 3
[$\ell(r_j)\leq3,\ \forall j$], then there is at most $2^n$ balanced
group presentations on $n$ relators, $\mv{X_n|R_n}$, having the same
abelianization.\eol
{\sc Proof.} As $\ell(r_i)\leq3$, commutators, $aba\inv b\inv$, can not
appear. Given 3 letters $a$, $b$ and $c$, there are only 2 inequivalent
relators having the abelianization $abc$:
\[
[abc]=[bca]=[cab] \hspace{3pc} [acb]=[cba]=[bac] \hspace{1pc}\Box
\]

Let us consider the (non-trivial) case when $\pi_1(C)$ has the trivial
abelianization: $H_1(C,Z)=0$. Balanced presentations of the trivial abelian
group are in correspondence with the unimodular integral
matrices\ftnt{In principle, $\det \wh{R}=\pm1$. We
choose +1 for convenience.}

\be
\wh{P}_E=\mv{x_1\ldots x_n|\sum_{j=1}^n \wh{R}_{ij} x_j} \hspace{2pc}
\wh{R}\in SL(n,Z)
\ee
We are interested in the subset corresponding to triangulations:

\be {\cal M}_n=\{\wh{R}\in SL(n,R): \sum_{j=1}^n \wh{R}_{ij}^2\leq3,\
\forall i\} 
\ee 

For all matrices $\wh{R}\in {\cal M}_n$, the allowed matrix elements are
$\wh{R}_{ij}=0,\pm1$, and there are at most 3 non-zero elements in each
row. We can independently permute rows and columns and multiply them by
$-1$.  It means the double coset $O(n,Z)\backslash {\cal M}_n/O(n,Z)$. The
group $O(n,Z)=S_n*Z_2$ is the free product of the symmetric group $S_n$ and
the group generated by the multiplication by the diagonal matrix
$diag(-1,+1,\ldots,+1)$.

An $\wh{R}$ matrix can be further reduced.  If the $i$'th row has only one
non-zero entry, $\wh{R}_{ij}\neq0$, then we can remove it along with the
$j$'th column by applying the AC2 and AC4 moves. If there are two non-zero
entries, $\wh{R}_{ij}\neq0$ and $\wh{R}_{ik}\neq0$, we nullify one of them
by using the AC2 move on columns. After that we can repeat the previous
step in order to decrease the size of the matrix.  Therefore, without loss
of generality, we assume that $(\wh{R}\wh{R}\tsp)_{jj}=3$
($j=1,\ldots,n$) and $(\wh{R}\tsp\wh{R})_{jj}\geq3$ ($j=1,\ldots,n$).

\noindent
{\sc Proposition~2.} Let a simplicial complex $C$ give (maybe after the 
described reduction) a presentation matrix such that
$(\wh{R}\wh{R}\tsp)_{jj}=3\ \forall j$.
If $C$ is a 3-manifold, then $\sum_{i=1}^n |\wh{R}_{ij}|=3$, $\forall j$.

\nnl {\sc Proof.} The reduced complex has only one 0-cell $\sigma^0$, $n$
1-cells $\sigma^1_k$ and $n$ 2-cells $\sigma^2_k$. Let us consider the
regular neighborhood of its 1-skeleton $K_1$ (cf.~\rf{Neuw}). The
neigborhood of $\sigma^0$ is a 3-ball $\beta^3$. Its boundary is the
2-sphere $\d \beta^3\cong S^2$. The neigborhoods of 1-cells $\sigma^1_k$
are solid tori. They intersect the sphere, $\d \beta^3$, in a disjoint
collection of disks. The 2-cells $\sigma^2_k$ intersect $\d \beta^3$ in a
number of simple arcs connecting the disks pairwise. The number of disks
equals $2n$, while that of arcs equals $3n$. Any such configuration defines
a graph in a natural way. If any edge $E$ of a graph connects distinct
vertices and the valency of each vertex $V$ is at least 3, then
$2\#E\geq3\#V$. As in the case in question $2\#E=3\#V=6n$, we conclude that
the valency of each vertex equals 3. $\Box$

Let us denote this restricted set of matrices as $\wh{{\cal M}}_n^3 \equiv
\{\wh{R} \in SL(n,Z)\ |\ (\wh{R}\tsp\wh{R})_{jj}=(\wh{R}\wh{R}\tsp)_{jj}=3\
\forall j\}$.  If $\wh{R}$ has a block form, $\det\wh{R}$ equals the product
of determinants of the blocks. Let us consider the irreducible case.  Any
matrix $\wh{R}\in\wh{{\cal M}}_n^3$ can be transformed into the form
$\wh{R}=I-B$. Then $B$ can be uniquely represented as the sum of two
$O(n,Z)$ matrices $B=P_1+P_2$: $P_{1,2}\in O(n,Z)$. This form has the
residual symmetry under conjugate permutations: $\wh{R}\to \sigma \wh{R}
\sigma^{-1}$, $\sigma\in O(n,Z)$. Therefore we can take the conjugacy
classes of the $B$ matrices as representatives for the double coset. For
example, we can specify the cycle structure of $P_1^{-1}P_2$ and an action
of $P_2$ on it. Therefore, if the constraint $\det\wh{R}=1$ is dropped, the
total number of configurations grows factorially with $n$. The growth rate
of the number of matrices giving $\det\wh{R}=0$ is factorial as well,
because it is sufficient to equate any two rows or columns to vanish the
determinant.

A convenient pictorial representation can be given in terms of oriented
graphs on $n$ vertices, ${\cal G}_n$. For it we interpret the $B$ matrices
as the adjacency matrices of the graphs.  If $B_{ij}\neq0$, we draw an
arrow from the $j$'th to the $i$'th vertex. There are exactly two links
going away from every vertex and exactly two arriving at it (one
corresponding to $P_1$ and the other to $P_2$). The determinant of $\wh{R}$
can be rewritten as the sum over all possible collections of oriented
self-avoiding loops $\{ L_1\ldots L_p\}$ in the graph ${\cal G}_n$

\be
\det(I-B)=\exp\Big(-\sum_{k=1}^{\infty}\frac{\tr B^k}k\Big)
= 1+\sum_{\{\scs L_1\ldots L_p\}} (-1)^p \prod_{k=1}^p 
\prod_{\scs (ji)\in L_k}B_{ij} 
\label{detR}
\ee
where $\prod_{(ji)\in L_k}$ goes over links constituting a loop $L_k$.

If we interpret $B$ as the incidence matrix (\ie if $\wh{R}_{ij}\neq0$,
the $i$'th edge is incident to the $j$'th vertex), then corresponding
graphs are always collections of disjoint circles. These give the cycle
structure of $P_1^{-1} P_2$.

Obviously $\det\wh{R}=1$, if the orgraph ${\cal G}_n$ has no closed
contours. It does not necessary mean that ${\cal G}_n$ is a tree, because
the loops in \eq{detR} are all oriented, \ie one is allowed to travel only
in directions shown by the arrows. If we had the stronger constraint,
$\sum_{j=1}^n |B_{ij}|\leq1$ (\ie exactly one arrow goes from every vertex),
then the only possible values of $\det\wh{R}$ would be 0, 1 and 2. In the
case $\det\wh{R}=1$, trees only would contribute. This simple model gives a
solution to the analogous counting problem for 2-dimensional spherical
triangulations (cf.~\rf{KKM}). 

Let us formulate the next intermediate result as

\nnl {\sc Proposition~3.} Let $|\wh{{\cal M}}_n^p|$ be the number of the
equivalence classes of matrices $\{ M \in SL(n,Z)\ |\ M' \cong M'',$ if
$M'=\omega_1 M'' \omega_2\ (\omega_{1,2} \in O(n,Z))\}$ obeying the
additional constraints $(M\tsp M)_{kk}=(MM\tsp)_{kk}=p$, $\forall k$. Then,
as $n \to \infty$ and $p$ finite, $|\wh{{\cal M}}_n^p| \le \exp(\lambda n)$
with some finite constant $\lambda$.

{\sc Proof} To prove that, we can use the following facts from the theory
of group lattices (see, for example, \rf{Raghu}): \\ (i) $SL(n,Z)$ is a
lattice in $SL(n,R)$, \ie $SL(n,R)/SL(n,Z)$ is connected and has a finite
volume with respect to any Haar measure on $SL(n,R)$.\\ (ii) A fundamental
domain of $SL(n,Z)$ in $SL(n,R)$ can be conveniently described in terms of
the Iwasawa decomposition: $M=\omega\alpha\eta$, where $\omega\in SO(n)$;
$\alpha = diag(\alpha_1,\alpha_2,\ldots, \alpha_n)$ (with all $\alpha_k>0$
and $\prod_{i=1}^n \alpha_i=1$); $\eta \in N$, where $N$ is the group of
upper-triangular unimodular matrices (\ie if $k>l$, $\eta_{kl}=0$; on the
diagonal $\eta_{kk}=1$; and, for $k<l$, $\eta_{kl}$ is arbitrary). A Siegel
set in $SL(n,R)$ is a set of the form $S_{tu}= \Omega A_t N_u$ defined by
the inequalities: $A_t=\{\alpha: \alpha_i \leq t\alpha_{i+1}\ (1\leq i <
n)\}$, $N_u=\{ \eta: |\eta_{kl}|\leq u\ (1\leq k < l \leq n)\}$ and
$\Omega$ coincides with the total $SO(n)$ group.  The fundamental domain is
the Siegel set $S_{tu}$ with $t=\frac2{\sqrt3}$ and $u=\frac12$. The Haar
measure in this parametrization is 

\be 
d\mu = d \omega
\prod_{i<j}\frac{\alpha_i} {\alpha_j}\prod_{i=1}^n d\alpha_i\;
\prod_{i<j} d\eta_{ij} \ee

\nnl
From the constraint
\be
\alpha_k^2 + \sum_{i<k}\alpha^2_i \eta^2_{ik} = p
\label{constrn}
\ee
we find that $\alpha_k^2 \leq p$ for all $k$. Together with the
unimodularity  property of matrices, it gives
\be
\prod_{j=2}^{n} (1 + \prod_{i<j} \eta^2_{ij}) \leq p^n. 
\label{cnst2}
\ee

\nnl From the observation that 

\be
(M\tsp M)_{ij}=\alpha_i^2\eta_{ij}+\sum_{k<i}\alpha^2_k\eta_{ki}\eta_{kj}
\hspace{1cm}   (i<j)
\ee

\nnl is a dilute matrix, it follows that $\eta_{nm}$ is dilute as well: in
any row of $\eta$ there are only a finite number of non-zero elements.
These equations mean that only a finite number of inequalities
$|\eta_{kl}| \leq \frac12$ can be violated for every value of $l$.

The action of $SL(n,Z)$ divides the $SL(n,R)$ group manifold into cells, each
of them identical to the fundamental domain. The equation
\be
\alpha_i=\frac2{\sqrt3}\alpha_{i+1}
\label{bndeq}
\ee
defines a hypersurface separating two different cells.

Now, we can use a standard trick in analytic number theory: to estimate the
number of distinct $SL(n,Z)$ matrices obeying some given inequalities, we can
simply count the number of cells inside a corresponding region in the
$SL(n,R)$ group manifold. 

As we are interested in counting the equivalence classes under
arbitrary permutations of rows and columns, let us assume in addition that

\be
\alpha_i\ge\alpha_{i+1}.
\label{ineq2}
\ee

\nnl For arbitrary $k \leq n$,

\be
\prod_{i=1}^k \alpha_i^2 = \det_{1\leq i,j \leq k} (M\tsp M)_{ij},
\ee

\nnl and the permutations of rows rearrange the $n$-tuples changing values of
$\alpha$'s, not simply permuting them. We can choose one in which the
inequalities(\ref{ineq2}) are satisfied. It is specific to our particular
case.

The unimodularity and the constraints (\ref{constrn}) together with the
inequalities (\ref{ineq2}) imply that, in the $n\to\infty$ limit, only a
finite number of equations (\ref{bndeq}) can be simultaneously
satisfied. If this number equals $k$, we can estimate the number of
distinct configurations as $(cn)^k/k!$ (with some constant $c$). As
upper-triangular unimodular matrices form a group over integers, we can
always obey the inequalities $|\eta_{ij}| \leq 1/2$ $\forall (i,j)$.
Analogously, for the left multiplications by $SO(n,Z)$ matrices, we can
choose any coordinates on the coset $SO(n,R)/SO(n,Z)$. Then, as follows
from the theory of group lattices, the integrality constraints on matrix
elements can have only one or no solution for any fixed tuple
$(\alpha_1,\alpha_2, \ldots ,\alpha_n)$.

The number of equivalence classes of dilute upper-triangular $\eta_{ij}$
matrices is exponentially bounded. The combinatorial reason for that is
that they have, in a general position, more symmetry than densely filled
matrices from the $N$ group. If $\eta_{nm}=0$ for $k\leq n,m \leq l$, then
$\sigma \eta \sigma\inv = \eta$ for all permutations $\sigma$ acting only
on the rows and columns from $k$ to $l$ and identical on the others
($\sigma \in I_{r'} \oplus S_q \oplus I_{r''}$, $q=l-k+1$). Let us take one
of such matrices and try to draw a consequence of this enhanced symmetry. We
have

\be
\sigma (\omega\alpha\eta) \sigma\inv = \omega' \alpha' \eta'
\ee

\nnl As $\omega'=\sigma\omega\sigma\inv\in SO(n)$ and
$\eta'=\sigma\eta\sigma\inv\in N$, $\alpha'=\sigma\alpha\sigma\inv$ has to
represent an integer point in the $\alpha$ space (\ie corresponding to a
matrix from $SL(n,Z)$). Let $\ell_{\rm min}$ be a minimal distance between
any integer points.  If $\alpha_i$ is permuted with $\alpha_j$ and
$|\log(\alpha_i/\alpha_j)|<<\ell_{\rm min}$, then, in a general position
inside a lattice cell, there must be $\alpha_i = \alpha_j$, because
otherwise there would be 2 different integer points in the fundamental
domain. Thus, we conclude that, in a general position, the product
$\alpha\eta$ possesses the same symmetry under permutations as the dilute
$\eta$ matrix alone. Then, in the $n\to \infty$ limit,
 the number of
equivalence classes of such products has to be of the same order as for the
$\eta$ matrices. $\Box$

From the pure mathematician's point of view, the argumentation given above
is not satisfactory.  Hopefully, a mathematically minded reader will be
able to complete the proof (or, maybe, find a loophole in it). Let us accept
the claim of this proposition as a plausible hypothesis. Then we can
conclude that

\nnl {\sc Theorem~1} The number of 3-dimensional simplicial manifolds
constructed of $n$ tetrahedra and having the trivial homology group,
$H_1(C,Z)=0$, grows at most exponentially with $n$: $|C_n| < e^{\lambda
n}$, for some finite constant $\lambda$.

\nnl {\sc Proof} To finish the proof, we have to show that, for any matrix
$M \in \wh{\cal M}_n^3$ from a given equivalence class, there are
exponentially many simplicial complexes accociated with it. In terms of
abelian presentations, the problem is reduced to estimating the number of
inequivalent multiplications by upper triangular $N_1\times(N_3+1)$
matrices. A multiplication by such a matrix from the left corresponds to a
change of the basis of relators in a $\pi_1$ presentation generated by AC2
and AC4 moves only. Let us remember that we are considering simplicial
complexes. These changes of bases corresponds to formal 2-deformations of
2-spines of complexes. Therefore their number is restricted from above by
the number of corresponding spanning trees in $\wt{K_1}\subset \wt{C}$.
As the valency of the trees is equal to 4, the statement follows.
$\Box$

\nnl
{\sc Theorem~2} The number of 3-dimensional simplicial manifolds constructed of
$n$ tetrahedra and having a fixed homology group $H_1(C,Z)$ grows at most
exponentially with $n$: $|C_n| < e^{\lambda n}$, for some finite constant
$\lambda$.

\nnl {\sc Proof} As was pointed out in Section~2, an arbitrary presentation
matrix can be transformed into a unique diagonal form by pre- and
post-multiplications by $SL(n,Z)$ matrices. Such multiplications correspond
to changes of basis of relators as well as generators. Therefore, the
statement is a consequence of the previous theorem. 
$\Box$

It is natural to assume that, having fixed the Betti number $b_1$ only, one
gets a factorial growth of the number of complexes as a function of a
volume. Indeed, we can see that the number of inequivalent presentation
matrices grows faster than an exponential in this case. On the other hand,
it is a piece of mathematical folklore that there is always at least one
3-manifold associated with a given abelian presentation\footnote{We
are indebted to V.Turaev for communicating this fact to us.}. Although,
such a manifold should not necessarily be simplicial, it is plausible
assumption that the simpliciality condition is not restrictive enough to
cut down the growth rate.


\newsection{Reduced model and connection with the 2-dimensional loop gas model}

As has been already discussed, in 3 dimensions there are several natural
choices of classes of complexes to define the partition function over. And
the question of universality classes is open. The simplest choice could be
the set of 3-spheres with imbedded 2-collapsible spines.  It looks very
natural from the viewpoint of Section~3. We restrict acceptable
presentations to the set of obviously trivial ones. Namely, to those that
can be reduced to the empty one, $\mv{|}$, by using only AC2 and AC4 moves
as follows from the little Wright's theorem. So, we arrive at the partition
function for the reduced model (cf. \eq{Z2} and (\ref{Z2b}))

\[
{\bf Z}^{(2)}_{\rm red}(\alpha,\mu) = \sum_{\{C \cong S^3\}}
\sum_{\{K_2 \subset C|K_2{\scs \stackrel{2}{\nearrow\!\searrow}}*\}}
\sum_{\{T\subset K_2\}} e^{\alpha N_0 - \mu N_3}  = 
\]\be
= \sum_{N_3,N_0} 
\sum_{\{P_{N_3+1;N_3+N_0}\cong\,\mv{|}\}} 
e^{\alpha N_0 - \mu N_3} 
\label{Z2red}
\ee

\nnl where $\sum_{\{C \cong S^3\}}$ is the sum over all simplicial
3-spheres; $\sum_{\{K_2 \subset C|K_2{\scs
\stackrel{2}{\nearrow\!\searrow}}*\}}$ is the sum over 2-collapsible spines,
$K_2$, embedded in $C$; $\sum_{\{T\subset K_2\}}$ is the sum over all
spanning trees in $K_2$. The abelianizations of all the presentations from
the sum $\sum_{\{P_{N_3+1;N_3+N_0}\cong\,\mv{|}\}}$ have upper-triangular
unimodular matrices.

Let us remind a reader that, by definition, a spine is a 2-skeleton $K_2$ of
a complex $C$ in a decomposition with only one 3-cell. In the simplicial
context, this 3-cell is a ball $B_3$ with a triangulated boundary. $K_2$ is
obtained by some pair-wise identification (gluings) of all boundary
triangles. If $K_2$ is 2-collapsible, one can use recursively only one
gluing operation: the identification of 2 triangles sharing a common link
on the boundary, in other words, a folding along the link. This can
simulate any sequence of elementary 2-collapses and expansions, with an
elementary 2-ball being just a single triangle.

In the dual language, one starts with an arbitrary spherical 3-valent fat
graph and applies the move which can be represented as the flip of a link
with the subsequent elimination of it:

\setlength{\unitlength}{1mm}
\be
\bp(80,20)(-40,-10) \thicklines \put(-35,0){\line(-1,1){8}}
\put(-35,0){\line(-1,-1){8}} \put(-35,0){\line(1,0){13}}
\put(-22,0){\line(1,-1){8}} \put(-22,0){\line(1,1){8}}
\put(-11,-5){\makebox(10,10){$\Longrightarrow$}} \put(5,5){\line(-1,1){5}}
\put(5,5){\line(1,1){5}} \put(5,-5){\line(-1,-1){5}}
\put(5,-5){\line(1,-1){5}} \multiput(5,-5)(0,1){10}{\line(0,1){0.5}}
\put(15,-5){\makebox(10,10){$\Longrightarrow$}}
\put(32,10){\oval(10,10)[b]} \put(32,-10){\oval(10,10)[t]} \ep
\label{move1}
\ee

\nnl Both of these moves are well known and have been used in Monte-Carlo
simulations of triangulated surfaces \cite{KKM}. Having glued all
triangles, one finishes with a collection of self-avoiding closed loops,
the number of which equals $N_0-1$. We weight every loop with
the numerical factor $e^{\alpha}$.

It is convenient, instead of erasing a links after a flip, to decorate it
with the dashed propagator and keep its track in a process of subsequent
foldings. To represent such configurations, we need to introduce the
infinite set of vertices

\be
\bp(130,20)(5,-10)
\thicklines
\multiput(10,0)(30,0){4}{\line(1,0){10}}
\multiput(15,0)(0,1){10}{\line(0,1){0.5}}
\multiput(45,0)(0,1){10}{\line(0,1){0.5}}
\multiput(75,0)(0,1){10}{\line(0,1){0.5}}
\multiput(105,0)(0,1){10}{\line(0,1){0.5}}
\multiput(45,0)(0,-1){10}{\line(0,1){0.5}}
\multiput(75,0)(-0.5,-1){10}{\line(0,-1){0.5}}
\multiput(75,0)(0.5,-1){10}{\line(0,-1){0.5}}
\multiput(105,0)(0,-1){10}{\line(0,1){0.5}}
\multiput(105,0)(-0.5,-1){10}{\line(0,-1){0.5}}
\multiput(105,0)(0.5,-1){10}{\line(0,-1){0.5}}
\put(125,-5){\makebox(10,10){and so on}}
\ep
\label{vert}
\ee

\noindent
which are generated by flips. For example,\\

\noindent
\bp(25,20)(-10,-10)
\thicklines
\put(-5,0){\line(1,0){10}}
\put(-5,0){\line(-1,-1){5}}
\put(-5,0){\line(-1,1){5}}
\put(5,0){\line(1,-1){5}}
\put(5,0){\line(1,1){5}}
\multiput(0,0)(0,-1){8}{\line(0,-1){0.5}}
\ep
\raisebox{9mm}{gives}
\bp(18,20)(-10,-10)
\thicklines
\put(-5,7){\line(1,0){10}}
\put(-5,0){\line(1,0){10}}
\multiput(0,0)(0,1){8}{\line(0,1){0.5}}
\multiput(0,0)(0,-1){8}{\line(0,-1){0.5}}
\ep
\raisebox{9mm}{\Large ;\hspace{5mm}}
\bp(25,20)(-10,-10)
\thicklines
\put(-5,0){\line(1,0){10}}
\put(-5,0){\line(-1,-1){5}}
\put(-5,0){\line(-1,1){5}}
\put(5,0){\line(1,-1){5}}
\put(5,0){\line(1,1){5}}
\multiput(-2,0)(0,-1){8}{\line(0,-1){0.5}}
\multiput(2,0)(0,-1){8}{\line(0,-1){0.5}}
\ep
\raisebox{9mm}{gives}
\bp(20,20)(-10,-10)
\thicklines
\put(-5,7){\line(1,0){10}}
\put(-5,0){\line(1,0){10}}
\multiput(0,0)(0,1){8}{\line(0,1){0.5}}
\multiput(0,0)(0.5,-1){8}{\line(0,-1){0.5}}
\multiput(0,0)(-0.5,-1){8}{\line(0,-1){0.5}}
\ep
\raisebox{9mm}{and so on} 

\noindent
\bp(22,20)(-10,-10)
\thicklines
\put(-5,0){\line(1,0){10}}
\put(-5,0){\line(-1,-1){5}}
\put(-5,0){\line(-1,1){5}}
\put(5,0){\line(1,-1){5}}
\put(5,0){\line(1,1){5}}
\multiput(-2,0)(0,-1){8}{\line(0,-1){0.5}}
\multiput(2,0)(0,1){8}{\line(0,-1){0.5}}
\ep
\raisebox{9mm}{or} 
\bp(22,20)(-10,-10)
\thicklines
\put(-5,0){\line(1,0){10}}
\put(-5,0){\line(-1,-1){5}}
\put(-5,0){\line(-1,1){5}}
\put(5,0){\line(1,-1){5}}
\put(5,0){\line(1,1){5}}
\multiput(-2,0)(0,1){8}{\line(0,-1){0.5}}
\multiput(2,0)(0,-1){8}{\line(0,-1){0.5}}
\ep
\raisebox{9mm}{or} 
\bp(22,20)(-10,-10)
\thicklines
\put(-5,0){\line(1,0){10}}
\put(-5,0){\line(-1,-1){5}}
\put(-5,0){\line(-1,1){5}}
\put(5,0){\line(1,-1){5}}
\put(5,0){\line(1,1){5}}
\multiput(0,0)(0,1){8}{\line(0,-1){0.5}}
\multiput(0,0)(0,-1){8}{\line(0,-1){0.5}}
\ep
\raisebox{9mm}{give}
\bp(20,20)(-10,-10)
\thicklines
\put(-5,3){\line(1,0){10}}
\put(-5,-3){\line(1,0){10}}
\multiput(0,-10)(0,1){20}{\line(0,1){0.5}}
\ep

\noindent
\bp(22,20)(-10,-10)
\thicklines
\put(-5,0){\line(1,0){10}}
\put(-5,0){\line(-1,-1){5}}
\put(-5,0){\line(-1,1){5}}
\put(5,0){\line(1,-1){5}}
\put(5,0){\line(1,1){5}}
\multiput(-3,0)(0,-1){8}{\line(0,-1){0.5}}
\multiput(0,0)(0,-1){8}{\line(0,-1){0.5}}
\multiput(3,0)(0,1){8}{\line(0,-1){0.5}}
\ep
\raisebox{9mm}{or} 
\bp(22,20)(-10,-10)
\thicklines
\put(-5,0){\line(1,0){10}}
\put(-5,0){\line(-1,-1){5}}
\put(-5,0){\line(-1,1){5}}
\put(5,0){\line(1,-1){5}}
\put(5,0){\line(1,1){5}}
\multiput(-3,0)(0,-1){8}{\line(0,-1){0.5}}
\multiput(0,0)(0,1){8}{\line(0,-1){0.5}}
\multiput(3,0)(0,-1){8}{\line(0,-1){0.5}}
\ep
\raisebox{9mm}{or} 
\bp(22,20)(-10,-10)
\thicklines
\put(-5,0){\line(1,0){10}}
\put(-5,0){\line(-1,-1){5}}
\put(-5,0){\line(-1,1){5}}
\put(5,0){\line(1,-1){5}}
\put(5,0){\line(1,1){5}}
\multiput(-3,0)(0,1){8}{\line(0,-1){0.5}}
\multiput(0,0)(0,-1){8}{\line(0,-1){0.5}}
\multiput(3,0)(0,-1){8}{\line(0,-1){0.5}}
\ep
\raisebox{9mm}{or} 
\bp(22,20)(-10,-10)
\thicklines
\put(-5,0){\line(1,0){10}}
\put(-5,0){\line(-1,-1){5}}
\put(-5,0){\line(-1,1){5}}
\put(5,0){\line(1,-1){5}}
\put(5,0){\line(1,1){5}}
\multiput(0,0)(-0.5,-1){8}{\line(0,-1){0.5}}
\multiput(0,0)(0.5,-1){8}{\line(0,-1){0.5}}
\multiput(0,0)(0,1){8}{\line(0,-1){0.5}}
\ep
\raisebox{9mm}{give} 
\bp(20,20)(-10,-8)
\thicklines
\put(-5,6){\line(1,0){10}}
\put(-5,0){\line(1,0){10}}
\multiput(0,0)(0,1){12}{\line(0,1){0.5}}
\multiput(0,0)(0.5,-1){8}{\line(0,-1){0.5}}
\multiput(0,0)(-0.5,-1){8}{\line(0,-1){0.5}}
\ep

\noindent
and so on.

In the end, we obtain planar diagrams with two types of propagators: solid
ones produce closed loops while dashed form arbitrary clusters. The total
number of vertices is equal to $2(N_3+1)$.

Different sequenses of flips can lead to differently looking planar diagrams 
corresponding to the same 3-dimensional complex. The simplest example is
given by the two possible self-energy reductions:

\be
\raisebox{-1.6cm}{
\setlength{\unitlength}{0.0004in}
\begin{picture}(4899,3324)(2000,-2923)
\thicklines
\put(4501,-2086){\oval(600,600)}
\put(4501,-436){\oval(600,600)}
\put(1801,-1561){\line( 1, 0){450}}
\put(151,-1561){\line( 1, 0){450}}
\put(1201,-1561){\oval(1200,750)}
\put(2401,-637){\vector( 2, 1){600}}
\put(3976,389){\line( 1, 0){1050}}
\put(2401,-2311){\vector( 2,-1){600}}
\put(3976,-2911){\line( 1, 0){1050}}
\multiput(4501,-2911)(0,200){3}{\line( 0, 1){100}}
\multiput(4501,-120)(0,200){3}{\line( 0, 1){100}}
\end{picture}}
\ee

\nnl It is easy to see that this overcounting exactly corresponds to taking
a sum over all spanning trees $\{T\subset K_2\}$ in \eq{Z2red}. More
precisely, links chosen to fold along are not in $T$.

Thus, the model can be represented as a non-local gas of self-avoiding
closed loops with the partition function

\be
Z^{(2)}_{\rm red}(\nu,\mu)=\sum_{\{{\cal D}\}} w_{\cal D}\;
                     e^{\alpha N_0-\mu N_3}
\label{nlocLG}
\ee

\nnl where $\sum_{\{{\cal D}\}}$ is the sum over all the diagrams. The
weight $w_{\cal D}$ is equal to the number of inequivalent starting
configurations giving the same planar diagram $\cal D$ (can be 0). The
number of closed loops is simply equal to $N_0-1$ and $\alpha$ can be
identified with the inverse Newton constant.

Let us take an arbitrary planar graph with dashed propagators. In order to
obtain all possible starting configurations, we have to flip back dashed
links in all possible ways. For any given dashed cluster, a result is a
subset of all planar 3-valent graphs. Hence, the number of them is
exponentially bounded as a function of the number of links. Let us give a
few illustrative examples. First, flips have to be performed in linear
clusters giving all possible planar 3-valent trees doubled in a
``mirror'':

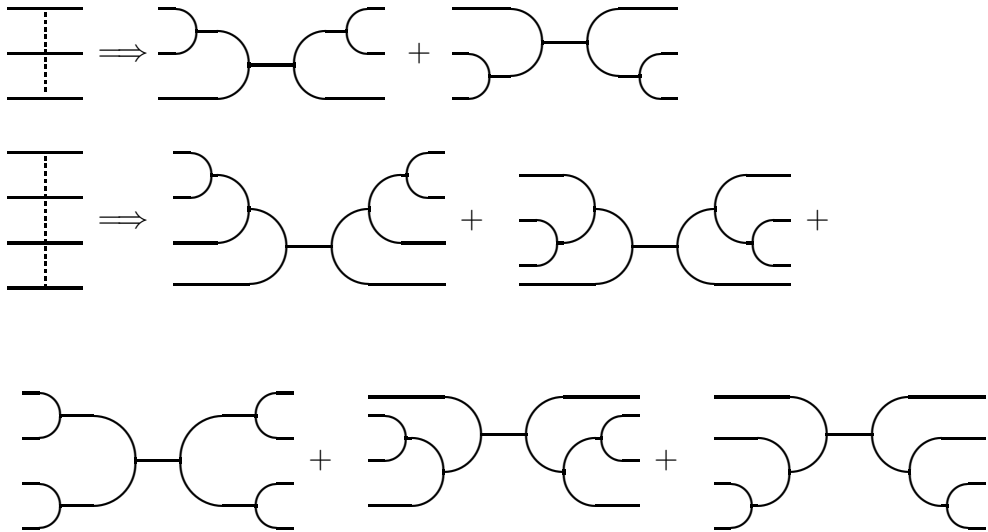
\bp(12,20)(-5,-10)
\thicklines
\put(-5,6){\line(1,0){10}}
\put(-5,0){\line(1,0){10}}
\put(-5,-6){\line(1,0){10}}
\multiput(0,0)(0,1){6}{\line(0,1){0.5}}
\multiput(0,0)(0,-1){6}{\line(0,1){0.5}}
\ep
\raisebox{9mm}{$\Longrightarrow$}
\bp(33,20)(-15,-10)
\thicklines
\put(-3,-1.5){\line(1,0){6}}
\put(10,-1.5){\oval(14,9)[l]}
\put(-10,-1.5){\oval(14,9)[r]}
\put(15,3){\oval(10,6)[l]}
\put(-15,3){\oval(10,6)[r]}
\put(-15,-6){\line(1,0){6}}
\put(15,-6){\line(-1,0){6}}
\ep
\raisebox{9mm}{+\ }
\bp(33,20)(-15,-10)
\thicklines
\put(-3,1.5){\line(1,0){6}}
\put(10,1.5){\oval(14,9)[l]}
\put(-10,1.5){\oval(14,9)[r]}
\put(15,-3){\oval(10,6)[l]}
\put(-15,-3){\oval(10,6)[r]}
\put(-15,6){\line(1,0){6}}
\put(15,6){\line(-1,0){6}}
\ep

\bp(12,20)(-5,-10)
\thicklines
\put(-5,9){\line(1,0){10}}
\put(-5,3){\line(1,0){10}}
\put(-5,-3){\line(1,0){10}}
\put(-5,-9){\line(1,0){10}}
\multiput(0,-9)(0,1){18}{\line(0,1){0.5}}
\ep
\raisebox{9mm}{$\Longrightarrow$}
\bp(40,20)(-20,-10)
\thicklines
\put(-3,-3.5){\line(1,0){6}}
\put(8,-3.5){\oval(10,10)[l]}
\put(-8,-3.5){\oval(10,10)[r]}
\put(8,-8.5){\line(1,0){10}}
\put(-8,-8.5){\line(-1,0){10}}
\put(13,1.5){\oval(10,9)[l]}
\put(-13,1.5){\oval(10,9)[r]}
\put(13,-3){\line(1,0){5}}
\put(-13,-3){\line(-1,0){5}}
\put(18,6){\oval(10,6)[l]}
\put(-18,6){\oval(10,6)[r]}
\ep
\raisebox{9mm}{+\ }
\bp(40,20)(-20,-10)
\thicklines
\put(-3,-3.5){\line(1,0){6}}
\put(8,-3.5){\oval(10,10)[l]}
\put(-8,-3.5){\oval(10,10)[r]}
\put(8,-8.5){\line(1,0){10}}
\put(-8,-8.5){\line(-1,0){10}}
\put(13,1.5){\oval(10,9)[l]}
\put(-13,1.5){\oval(10,9)[r]}
\put(13,6){\line(1,0){5}}
\put(-13,6){\line(-1,0){5}}
\put(18,-3){\oval(10,6)[l]}
\put(-18,-3){\oval(10,6)[r]}
\ep
\raisebox{9mm}{+\ }
\eol

\bp(40,20)(-20,-10)
\thicklines
\put(-3,0){\line(1,0){6}}
\put(13,0){\oval(20,12)[l]}
\put(-13,0){\oval(20,12)[r]}
\put(18,6){\oval(10,6)[l]}
\put(-18,6){\oval(10,6)[r]}
\put(18,-6){\oval(10,6)[l]}
\put(-18,-6){\oval(10,6)[r]}
\ep
\raisebox{9mm}{+\ }
\bp(40,20)(-20,-10)
\thicklines
\put(-3,3.5){\line(1,0){6}}
\put(8,3.5){\oval(10,10)[l]}
\put(-8,3.5){\oval(10,10)[r]}
\put(8,8.5){\line(1,0){10}}
\put(-8,8.5){\line(-1,0){10}}
\put(13,-1.5){\oval(10,9)[l]}
\put(-13,-1.5){\oval(10,9)[r]}
\put(13,-6){\line(1,0){5}}
\put(-13,-6){\line(-1,0){5}}
\put(18,3){\oval(10,6)[l]}
\put(-18,3){\oval(10,6)[r]}
\ep
\raisebox{9mm}{+\ }
\bp(40,20)(-20,-10)
\thicklines
\put(-3,3.5){\line(1,0){6}}
\put(8,3.5){\oval(10,10)[l]}
\put(-8,3.5){\oval(10,10)[r]}
\put(8,8.5){\line(1,0){10}}
\put(-8,8.5){\line(-1,0){10}}
\put(13,-1.5){\oval(10,9)[l]}
\put(-13,-1.5){\oval(10,9)[r]}
\put(13,3){\line(1,0){5}}
\put(-13,3){\line(-1,0){5}}
\put(18,-6){\oval(10,6)[l]}
\put(-18,-6){\oval(10,6)[r]}
\ep

\noindent
and so on. The number of initial configurations for a linear claster of $n$
dashed links, $C_n$, is given by Catalan's numbers generated by the
function

\be
C(x)=\sum_{n=0}^{\infty}C_nx^n:=\sum_{n=0}^{\infty}\
\raisebox{-1cm}{$
\overbrace{
\bp(10,20)(-5,-10)
\thicklines
\put(0,0){\circle{10}}
\put(-1.5,-1.5){\mbox{\sf\bf T}}
\put(0,5){\line(0,1){5}}
\put(1,5){\line(1,3){1.66}}
\put(2,4.5){\line(2,3){3.33}}
\put(-1,5){\line(-1,3){1.66}}
\put(-2,4.5){\line(-2,3){3.33}}
\put(0,-5){\line(0,-1){5}}
\ep
}^{n+1}$}\
x^n=\frac1{2x}(1-\sqrt{1-4x})
\ee

After that dashed propagators attached to the ends of the 
linear cluster have to be 
reattached in all possible ways to links of these ``doubled trees''. The 
number of configurations can be estimated as follows.
First, we have to calculate the generating function for the number of 
possible attachments of dashed propagators to a solid line:

\be
q(\nu,\mu):=\sum_{n,m}\
\raisebox{-1cm}{$
\overbrace{\underbrace{
\bp(20,20)(-10,-10)
\thicklines
\put(0,0){\circle*{3}}
\put(-8,0){\line(1,0){16}}
\multiput(0,0)(0,1){10}{\line(0,1){0.5}}
\multiput(0,0)(0.5,1){9}{\line(0,1){0.5}}
\multiput(0,0)(0.75,0.75){9}{\line(0,1){0.5}}
\multiput(0,0)(-0.5,1){9}{\line(0,1){0.5}}
\multiput(0,0)(-0.75,0.75){9}{\line(0,1){0.5}}
\multiput(0,0)(0,-1){10}{\line(0,-1){0.5}}
\multiput(0,0)(-0.5,-1){9}{\line(0,-1){0.5}}
\multiput(0,0)(-0.75,-0.75){9}{\line(0,-1){0.5}}
\multiput(0,0)(0.5,-1){9}{\line(0,-1){0.5}}
\multiput(0,0)(0.75,-0.75){9}{\line(0,-1){0.5}}
\ep
}_m
}^{n}$}\
\nu^n\mu^m
\ee
It can be obtained from the equation

\be
q(\nu,\mu)=\Big[\frac{\nu}{1-\mu}+\frac{\mu}{1-\nu}-\nu\mu\Big] 
q(\nu,\mu)+1
\ee
which gives

\be
q(\nu,\mu)=\frac{(1-\nu)(1-\mu)}{(1-\nu-\mu)^2-(\nu+\mu)\nu\mu 
+\nu^2\mu^2}
\ee

Second, we have to know the generating function for the number of all 
possible attachments of dashed propagators to a $k$-legged tree on the 
left (or on the right but not on both sides):

\be
t(x,\nu):=\sum_{k,n} \raisebox{-1mm}{${\scr n}\Big\{$}\hspace{5mm}
\raisebox{-1cm}{$
\overbrace{
\bp(10,20)(-5,-10)
\thicklines
\put(0,0){\circle{10}}
\put(-1.5,-1.5){\mbox{\sf\bf T}}
\put(0,5){\line(0,1){5}}
\put(1,5){\line(1,3){1.66}}
\put(2,4.5){\line(2,3){3.33}}
\put(-1,5){\line(-1,3){1.66}}
\put(-2,4.5){\line(-2,3){3.33}}
\put(0,-5){\line(0,-1){5}}
\multiput(-5,0)(-1,0){5}{\line(-1,0){0.5}}
\multiput(-5,0.5)(-1,0.25){5}{\line(-1,0){0.5}}
\multiput(-5,1)(-1,0.5){5}{\line(-1,0){0.5}}
\multiput(-5,-0.5)(-1,-0.25){5}{\line(-1,0){0.5}}
\multiput(-5,-1)(-1,-0.5){5}{\line(-1,0){0.5}}
\ep
}^{k+1}$}\
x^{k}\nu^n;
\hspace{2pc} t(x,0)=C(x)
\ee
The equation for it is simply

\be
t(x,\nu)=xC(x)\frac{t(x,\nu)}{1-\nu}+1
\ee
which gives

\be
t(x,\nu)=\frac{1-\nu}{1-\nu-xC(x)}
\ee
Then, the quantity we need to calculate is

\[
T(x,\nu,\mu)=\sum_{n,m,k}
\raisebox{-1mm}{${\scr n}\Big\{$}\hspace{5mm}
\raisebox{-5mm}{$
\underbrace{
\bp(20,10)(-10,-5)
\thicklines
\multiput(-10,0)(1,0){20}{\line(1,0){0.5}}
\multiput(-10,-5)(5,0){5}{\line(0,1){10}}
\multiput(-10,0)(-1,0){5}{\line(-1,0){0.5}}
\multiput(-10,0.5)(-1,0.25){5}{\line(-1,0){0.5}}
\multiput(-10,1)(-1,0.5){5}{\line(-1,0){0.5}}
\multiput(-10,-0.5)(-1,-0.25){5}{\line(-1,0){0.5}}
\multiput(-10,-1)(-1,-0.5){5}{\line(-1,0){0.5}}
\multiput(10,0)(1,0){5}{\line(1,0){0.5}}
\multiput(10,0.5)(1,0.25){5}{\line(1,0){0.5}}
\multiput(10,1)(1,0.5){5}{\line(1,0){0.5}}
\multiput(10,-0.5)(1,-0.25){5}{\line(1,0){0.5}}
\multiput(10,-1)(1,-0.5){5}{\line(1,0){0.5}}
\ep
}_{k+1}$}\hspace{5mm}
\raisebox{-1mm}{$\Big\}{\scr m}$}
\; x^{k}\nu^n\mu^m
=\sum_{n,m,k}\raisebox{-3cm}{
\bp(30,60)(-15,-30)
\thicklines
\put(0,-5){\line(0,1){10}}
\put(0,0){\circle*{3}}
\multiput(0,0)(-1,0){8}{\line(-1,0){0.5}}
\multiput(0,0.5)(-1,0.25){7}{\line(-1,0){0.5}}
\multiput(0,1)(-1,0.5){6}{\line(-1,0){0.5}}
\multiput(0,-0.5)(-1,-0.25){7}{\line(-1,0){0.5}}
\multiput(0,-1)(-1,-0.5){6}{\line(-1,0){0.5}}
\multiput(0,0)(1,0){8}{\line(1,0){0.5}}
\multiput(0,0.5)(1,0.25){7}{\line(1,0){0.5}}
\multiput(0,1)(1,0.5){6}{\line(1,0){0.5}}
\multiput(0,-0.5)(1,-0.25){7}{\line(1,0){0.5}}
\multiput(0,-1)(1,-0.5){6}{\line(1,0){0.5}}
\put(0,15){\oval(20,20)[b]}
\put(-10,20){\circle{10}}
\put(-11.5,18.5){\mbox{\sf\bf T$\mbox{}_1$}}
\put(-10,25){\line(0,1){5}}
\put(-9,25){\line(1,3){1.66}}
\put(-8,24.5){\line(2,3){3.33}}
\put(-11,25){\line(-1,3){1.66}}
\put(-12,24.5){\line(-2,3){3.33}}
\multiput(-15,20)(-1,0){5}{\line(-1,0){0.5}}
\multiput(-15,20.5)(-1,0.25){5}{\line(-1,0){0.5}}
\multiput(-15,21)(-1,0.5){5}{\line(-1,0){0.5}}
\multiput(-15,19.5)(-1,-0.25){5}{\line(-1,0){0.5}}
\multiput(-15,19)(-1,-0.5){5}{\line(-1,0){0.5}}
\put(10,20){\circle{10}}
\put(8.5,18.5){\mbox{\sf\bf T$\mbox{}_2$}}
\put(10,25){\line(0,1){5}}
\put(9,25){\line(-1,3){1.66}}
\put(8,24.5){\line(-2,3){3.33}}
\put(11,25){\line(1,3){1.66}}
\put(12,24.5){\line(2,3){3.33}}
\multiput(15,20)(1,0){5}{\line(1,0){0.5}}
\multiput(15,20.5)(1,0.25){5}{\line(1,0){0.5}}
\multiput(15,21)(1,0.5){5}{\line(1,0){0.5}}
\multiput(15,19.5)(1,-0.25){5}{\line(1,0){0.5}}
\multiput(15,19)(1,-0.5){5}{\line(1,0){0.5}}
\put(0,-15){\oval(20,20)[t]}
\put(-10,-20){\circle{10}}
\put(-11.5,-21.5){\mbox{\sf\bf T$\mbox{}_1$}}
\put(-10,-25){\line(0,-1){5}}
\put(-9,-25){\line(1,-3){1.66}}
\put(-8,-24.5){\line(2,-3){3.33}}
\put(-11,-25){\line(-1,-3){1.66}}
\put(-12,-24.5){\line(-2,-3){3.33}}
\multiput(-15,-20)(-1,0){5}{\line(-1,0){0.5}}
\multiput(-15,-20.5)(-1,-0.25){5}{\line(-1,0){0.5}}
\multiput(-15,-21)(-1,-0.5){5}{\line(-1,0){0.5}}
\multiput(-15,-19.5)(-1,0.25){5}{\line(-1,0){0.5}}
\multiput(-15,-19)(-1,0.5){5}{\line(-1,0){0.5}}
\put(10,-20){\circle{10}}
\put(8.5,-21.5){\mbox{\sf\bf T$\mbox{}_2$}}
\put(10,-25){\line(0,-1){5}}
\put(9,-25){\line(-1,-3){1.66}}
\put(8,-24.5){\line(-2,-3){3.33}}
\put(11,-25){\line(1,-3){1.66}}
\put(12,-24.5){\line(2,-3){3.33}}
\multiput(15,-20)(1,0){5}{\line(1,0){0.5}}
\multiput(15,-20.5)(1,-0.25){5}{\line(1,0){0.5}}
\multiput(15,-21)(1,-0.5){5}{\line(1,0){0.5}}
\multiput(15,-19.5)(1,0.25){5}{\line(1,0){0.5}}
\multiput(15,-19)(1,0.5){5}{\line(1,0){0.5}}
\ep}
x^{k}\nu^n\mu^m 
\]
\be
=q(\nu,\mu)\Bigg\{1+\frac{x(1-\nu)^2(1-\mu)^2}
{\Big((1-\nu)^2-xC(x)\Big)\Big((1-\mu)^2-xC(x)\Big)}\Bigg\}
\label{T}
\ee

Newly appearing clusters have to be expanded in all possible ways 
according to the same procedure. What is important is that their number 
is restricted by the coefficients of $q(\nu,\mu)$, which is a factor in 
(\ref{T}), and does not depend on a volume of the trees. Therefore, one 
simply iterates till all dashed links disappear. 
The combinatorial reason for the existence of 
an exponential bound  for the number of obtained configurations is 
clear: they are all planar! 

We have just seen that the weights $w_{\cal D}$ in \eq{nlocLG} are
recursively calculable. However, it is difficult to really take account of
the entropy of configurations and the model (\ref{nlocLG}) seems to be
analyticly unsolvable. Therefore, as was first proposed in \rf{Bo}, let us
consider the localized version of it, namely, the dense phase of the
self-avoiding loop gas matrix model \cite{O(n)} (in the sequel simply loop
gas (LG) model):

\[
Z_{LG} =\int d^{\scs N^2}Y d^{\scs N^2}\prod_{\nu=1}^n X_{\nu} 
\exp\Big[-\frac{N}2\tr Y^2-\frac{N}2\sum_{\nu=1}^{n}\tr X^2_{\nu} 
+\frac{\lambda N}2\sum_{\nu=1}^{n}\tr YX^2_{\nu}\Big]
\]\be
=\int d^{\scs N^2}Y
\exp\Big[-\frac{N}2\tr Y^2 -\frac{n}2\tr\mbox{}^2\log\Big(1\otimes 1
- \lambda(Y\otimes 1 + 1 \otimes Y) \Big) \Big]
\label{LG}
\ee
Here $Y$ and $X_{\nu}$ are hermitian $N\times N$ matrices.
We have attached the lower index to the gaussian $X_{\nu}$ variable 
to weight every closed loop with the factor $n = e^{\alpha}$. 
$\lambda = e^{-\mu/2}$. 
In the $N\to\infty$ limit, this model generates planar diagrams having only
the simplest 3-valent vertices from the whole series (\ref{vert}). For such
diagramms, all the weights in \eq{nlocLG} are trivial: $w_{\cal D}=1$.

Let us suppose that this truncated model bears some qualitative features of
3-dimensional simplicial gravity. It seems to be a reasonable assumption,
because, as follows from the results of \rf{O(n)}, the corresponding
universality class is very large and includes all local perturbations of
the model (\ref{LG}). On the other hand, multiple overlapping sequences of
the flips can be regarded as a kind of renormalization group
procedure, which could draw any given system to a stable fixed point in some
imaginable space of all planar-graph models. It looks a bit more natural in
dual terms. One takes an arbitrary spherical ball with a huge triangulated
boundary and uses subsequently and randomly the folding operation (dual to
the flip). Soon the boundary triangulation will be randomized. If this
randomization has appropriate statistical properties, we could
expect to find many models falling in the same universality
class and, therefore, having the same continuum limit.

The critical behavior of the loop gas matrix model is well known.  For
$n<2$, it describes 2-d gravity interacting with $c<1$ conformal matter

\be
c=1-6\frac{(1-g_0)^2}{g_0}; \hspace{2pc} n=-2\cos \pi g_0
\ee
while for $n>2$ the 
corresponding matter is non-critical and the model trivializes.
In this phase, the number of closed loops is proportional to the volume 
and the mean length of each remains finite.

In the vicinity of a critical point $\mu_c$, the partition function 
behaves as

\be 
Z_{LG} \approx (\mu_c-\mu)^{2-\gamma_{\rm str}} 
\label{defgam}
\ee 
This formula
defines the famous string susceptibility exponent $\gamma_{\rm str}$.

Two main quantities of interest are the mean lattice volume

\be
\mv{N_3}_{LG}
=\frac{\d\ }{\d\mu} \log Z_{LG} \approx -(2-\gamma_{\rm str}) 
\frac{\mu_c}{\mu_c-\mu}
\ee
and the mean lattice curvature
\[
\mv{N_0}_{LG}
=\frac{\d\ }{\d\alpha} \log Z_{\scs LG} \approx (2-\gamma_{\rm str})
\frac{n\mu'_c(n)}{\mu_c(n)-\mu} - \frac{\d\gamma_{\rm str}} 
{\d\alpha}\log|\mu_c-\mu|
\]\be
\approx-\frac{\mu'_c}{\mu_c}\mv{N_3}_{LG} +\frac{\d\gamma_{\rm str}} 
{\d\alpha}\log\mv{N_3}_{LG}
\ee
where the coefficients can be calculated explicitly using the Gaudin and 
Kostov's exact solution \cite{O(n)}

\be
\frac{\d\ }{\d n}\log\mu_c= -\frac12\frac1{n+2};
\hspace{2pc}
\frac{\d\gamma_{\rm str}}{\d n}=\frac1{2\pi g_0^2 \sin\pi(1-g_0)}
\ee
If $2-n\ll 1$, then \cite{Bo}

\be
\mv{N_0}_{LG} \approx \frac14\mv{N_3}_{LG} 
+\frac1{\pi\sqrt{2-n}}\log\mv{N_3}_{LG}
\label{latcurv}
\ee

If $n>2$, then $\gamma_{\rm str}$ is independent of $n$ and 
$\mv{N_0}_{LG}$ is proportional to $\mv{N_3}_{LG}$.

This behavior is strikingly similar to the results of numerical
simulations. In \rf{3dsim2}, a phase transition in the model (\ref{Z}) with
respect to the Newton coupling, $\alpha$, was found. In
the ``elongated'' phase, $\alpha>\alpha_c$, the mean number of vertices,
$\mv{N_0}$, is strictly proportional to a volume, $\mv{N_3}$. In the
``crumpled'' phase, $\alpha<\alpha_c$, $\mv{N_0}$ is a non-trivial function of
$\mv{N_3}$. The analogy with the loop gas matrix model suggests that the
most probable scaling is $\mv{N_0}\approx c_1\mv{N_3} +c_2\log\mv{N_3}$
with $c_2$ singular at the critical point $\alpha_c$.  Presumably, it is a
type of this singularity that can be, in principle, calculated in continuum
theory in order to compare predictions of both approaches.  The obvious
problem for such a hypothetical comparison is that $\alpha$ is a bare
coupling. Therefore, in the lattice model, only critical points with
respect to it could show some universal features. In any case, \eq{latcurv}
looks reasonable. For the dimensionful mean curvature, $\mv{R}=2\pi a
\mv{N_0}_{LG}+ca\mv{N_3}_{LG}$, as a function of the voulume, $V=a^3
\mv{N_3}_{LG}$, we find

\be
\mv{R} \approx \frac{c_1}{a^2} V
+\frac{2a}{\sqrt{2-n}}\log V+\ldots
\label{curscal}
\ee

\nnl As the scalar curvature has to undergo an additive renormalization, it
is natural to find a linear volume term singular in the continuum limit:
$a\to0$, $V$ finite. The logarithmic in volume term can have in this limit
a finite coefficient, if simultaneously $n \to  2 - 0(a^2)$. This is quite
physical behavior. 

Of course, the loop gas matrix model cannot give us precise quantitative
information about simplicial gravity. However, it is very plausible that it
has qualitatively the same phase structure as models (\ref{Z}) and
(\ref{Z2}) and may be quite instructive from this point of view.

\newsection{Discussion}

\begin{enumerate}

\item Combinatorial group theory gives a natural mathematical framework and
sets up a standard language for physical problems related to lattice models
of 3-dimensional quantum gravity. All the formal group constructions with
relators and generators have a natural geometrical realization in terms of
2-dimensional complexes (or fake surfaces, in a less formal parlance).  And
vise versa, geometrical constructions can be formalized in the group theory
terms. It would be interesting to find physical models
which could be formulated and solved entirely in terms of abelian
presentations. It might be a mathematically adequate way to make physically
meaninful approximations. 

\item To determine a growth rate for different classes of complexes is the
first necessary step in the process of constructing and investigating
simplicial gravity models. It is an interesting field of research in its
own right, altough more mathematical in spirit. Explicit forms of
the asymptotic expansions would give rigorous and exact solutions of the
correponding physical models.

\item The status of the reduced model (\ref{Z2red}) is not yet absolutely
clear. As it was formulated in \eq{Z2red}, it looks as a genuine simplicial
gravity model with some peculiar matter coupled to it.  It is not clear if
this matter has any local description in terms of fields in a manifold. The
first natural question about the model to ask is if all simplicial spheres
are really counted in ${\bf Z}^{(2)}_{\rm red}(\alpha,\mu)$. Or, in other
words, whether any simplicial 3-sphere allows for embedding at least one
2-collapsible spine in it. At first thought, Zeeman's Dunce Hat could serve
as a source of counterexamples. However, triangles lying inside an initial
ball can form such a fake surface. Even in the simplest case of 2
tetrahedra a Zeeman's Dunce Hat can easily be obtained. And, say, the
complex constructed in \rf{DJ} cannot be a counterexample because it is not
simplicial. It seems to be an uneasy question in general. Say, the similar
statement in case of simply-connected manifolds would automaticly prove the
famous Poincar\'{e} hypothesis.

\item The model (\ref{Z2red}) can be successfully simulated numerically
with help of the Monte Carlo technique. Results of the simulations as well
as an account of the algorithm will be reported elsewhere \cite{ABKW}.  Let
us simply mention here that they support the hypothesis that both the
reduced 3-dimensional model (\ref{Z2red}) and the 2-dimensional loop gas
model belong to the same universality class in the continuum limit, with
the Newton coupling playing a role analogous to the central charge in 2D
gravity. It would be extremely interesting if the original model (\ref{Z2})
had a similar continuum limit. In principle, it would  not contradict to what
is known from numerical simulations of the pure 3D gravity model. If we
remember that the analogous 2-dimensional matrix model discribes the $c=-2$
matter coupled to gravity, which is topological, then it might seem less
surprising that our 3-dimensional construction is equivalent to some
2-dimensional model.

\end{enumerate}

\bigskip

{\Large\bf Acknowledgments}

\bigskip

I would like to thank J.Ambj\o rn, B.Durhuus, V.Kazakov, N.Kawamoto,
V.Turaev and Y.Watabiki for the fruitful discussions. This work has been
partially supported by the grants RFFI-97-02-17927 and INTAS-96-0524.

\end{document}